\documentclass[a4paper,conference]{IEEEtran}
\usepackage[left=0.514in,right=0.514in,top=0.75in,bottom=1.57in]{geometry}
\usepackage{hyperref}
\usepackage{graphicx}
\usepackage{placeins}
\usepackage{tabularx}
\usepackage{xcolor}
\usepackage{caption}
\usepackage{subcaption}

\ifCLASSINFOpdf
\else
\fi
\begin{document}
%
\title{Influence of Event Duration on Automatic Wheeze Classification}

\author{\IEEEauthorblockN{Bruno M. Rocha}
\IEEEauthorblockA{Univ Coimbra\\CISUC\\DEI\\Coimbra, Portugal\\bmrocha@dei.uc.pt}
\and
\IEEEauthorblockN{Diogo Pessoa}
\IEEEauthorblockA{Univ Coimbra\\CISUC\\DEI\\Coimbra, Portugal\\dpessoa@dei.uc.pt}
\and
\IEEEauthorblockN{Alda Marques}
\IEEEauthorblockA{Univ Aveiro\\Lab3R\\ESSUA, iBiMED\\Aveiro, Portugal\\amarques@ua.pt}
\and
\IEEEauthorblockN{Paulo Carvalho}
\IEEEauthorblockA{Univ Coimbra\\CISUC\\DEI\\Coimbra, Portugal\\carvalho@dei.uc.pt}
\and
\IEEEauthorblockN{Rui Pedro Paiva}
\IEEEauthorblockA{Univ Coimbra\\CISUC\\DEI\\
Coimbra, Portugal\\ruipedro@dei.uc.pt}
}


%


\maketitle

\begin{abstract}
Patients with respiratory conditions typically exhibit adventitious respiratory sounds, such as wheezes. Wheeze events have variable duration. In this work we studied the influence of event duration on wheeze classification, namely how the creation of the non-wheeze class affected the classifiers' performance. First, we evaluated several classifiers on an open access respiratory sound database, with the best one reaching sensitivity and specificity values of 98\% and 95\%, respectively. Then, by changing one parameter in the design of the non-wheeze class, i.e., event duration, the best classifier only reached sensitivity and specificity values of 55\% and 76\%, respectively. These results demonstrate the importance of experimental design on the assessment of wheeze classification algorithms' performance.
\end{abstract}

\renewcommand\IEEEkeywordsname{Keywords}

\section{Introduction}\label{sec:introduction}
Respiratory conditions are leading causes of morbidity and mortality worldwide \cite{WHO2018} and generate a significant burden for public health systems and society \cite{Gibson2013}. Early diagnosis and routine monitoring of patients with respiratory conditions are important for timely interventions; thus, considerable research efforts have been focused on these issues in the last couple of decades \cite{Marques2014}.

Respiratory sounds are a simple, objective, and noninvasive marker to assess patients’ respiratory condition \cite{jacome2015}. In clinical practice they are commonly assessed with pulmonary auscultation using a stethoscope. The expert clinician is trained to listen to and to recognize pathological findings, such as the presence of adventitious sounds (e.g. crackles, wheezes), commonly in the anterior and posterior chest of the patient \cite{Fleming2011}. Conventional auscultation is commonly applied but has some drawbacks, such as the inability to provide continuous monitoring and its inter-listener variability. Automated respiratory sound analysis could potentially overcome these limitations.

Respiratory sounds are divided into normal and abnormal categories. Normal respiratory sounds are nonmusical sounds produced from breathing and heard over the trachea and chest wall \cite{Sovijarvi2000}. They show different acoustic properties, such as duration, pitch, and sound quality depending on subjects’ characteristics, subjects’ position, respiratory flow, and place of recording \cite{Pramono2017a,Oliveira2014}. On the other hand, adventitious sounds are abnormal sounds that are superimposed on the normal respiratory sounds \cite{Sovijarvi2000,Bardou2018}. Adventitious sounds can be categorized in two main types, continuous and discontinuous \cite{Hadjileontiadis2018}. The nomenclature preconized by the standardization of lung sound nomenclature taskforce \cite{Pasterkamp2016}, i.e., continuous adventitious sounds are referred to as wheezes, while discontinuous adventitious sounds are called crackles, will be followed in this study.

Crackles are explosive, short, discontinuous, and nonmusical adventitious sounds that are attributed to the sudden opening and closing of abnormally closed airways \cite{marques2018_10}. In contrast, wheezes are musical respiratory sounds usually longer than 80-100 ms and with frequencies ranging from less than 100 Hz to 1000 Hz, with harmonics that occasionally exceed 1000 Hz \cite{Bohadana2014a}. Wheezes occur when there is a flow limitation and they can be produced by all mechanisms that reduce airway caliber. Clinically, they can be defined by their frequency (mono- or polyphonic), intensity, number, duration, and position in the respiratory cycle (inspiratory or expiratory), gravity influence, and respiratory maneuvers \cite{marques2018_10}. Wheezes have been used for diagnostic purposes in several respiratory conditions in children (e.g., bronchiolitis) and in adults (e.g., chronic obstructive pulmonary disease (COPD)) \cite{marques2018_10}.

In this work we propose several machine learning methods for the classification of wheeze events against other types of events, randomly selected. The main objectives of this study were to: a) develop several wheeze classification models (traditional and deep learning approaches) and b) study the influence of the randomly generated events on the classification performance.

The article is organized as follows: in \autoref{sec:background}, we provide a general overview of several state-of-the-art algorithms that have been used in similar works; in \autoref{sec:methods}, we present the database, as well as showing all the methods used in the different experiments; in \autoref{sec:evaluation}, the obtained results are displayed; and, lastly, we discuss the results and give a conclusion in \autoref{sec:discussion}.

\section{Related Work}\label{sec:background}

\begin{table*}[t!]
\caption{Summary of selected works}
\label{tab:SOTA}
\resizebox{\textwidth}{!}{%
\begin{tabular}{c|c|c|c}
\hline
\textbf{Ref} & \textbf{Data} & \textbf{Classes} & \textbf{Best Results} \\\hline
\cite{Forkheim1995} & Prt: NA; Rec: NA; Src: Priv & 2: Wheezes and Normal & Acc: 96\% \\
\cite{Riella2009} & Prt: NA ; Rec: 28 Src: R.A.L.E. & 2: Wheezes and Normal & Acc: 85\%; Sens: 86\% ; Spec: 82\% \\
\cite{Chamberlain2016} & Prt: 284; Rec: 500; Src: Priv & 3: Wheezes, Crackles, and Normal & Wheeze AUC: 86\% \\
\cite{Lozano2016a} & Prt: 30; Rec: 870; Src: Priv & 2: Wheezes and Normal & Acc: 94\%; Prec: 95\%; Sens: 94\%; Spec: 94\% \\
\cite{Mendes2015} & Prt: 12; Rec: 24; Src: Priv & 2: Wheezes and Normal & Acc: 98\%; Sens: 91\%; Spec: 99\%; MCC: 93\% \\
\cite{Aykanat2017} & Prt: 1630; Rec: 17930; Src: Priv & 2: Healthy and Pathological & Acc: 86\%; Prec: 86\%; Sens: 86\%; Spec: 86\% \\
\cite{Bardou2018} & Prt: 15; Rec: 15; Src: R.A.L.E. & 7, including Wheezes, Crackles, and Normal & Acc: 96\%; Wheeze Prec: 98\%; Wheeze Sens: 100\% \\
\cite{Serbes2018} & Prt: 126; Rec: 920; Src: RSD & 3: Wheezes, Crackles, and Normal & Wheeze Sens: 79\%; Spec: 91\% \\
\cite{Jakovljevic2018} & Prt: 126; Rec: 920; Src: RSD & 3: Wheezes, Crackles, and Normal & Wheeze Sens: 52\%; Spec: 52\% \\
\cite{Kochetov2018} & Prt: 126; Rec: 920; Src: RSD & 4: Wheezes, Crackles, Wheezes+Crackles, and Normal & Wheeze Sens: 70\%; Spec: 74\% \\
\cite{Chen2019} & Prt: NA; Rec: 240; Src: R.A.L.E. and RSD & 2: Wheezes and Normal & Acc: 99\%; Sens: 96\%; Spec: 99\% \\
\cite{Ntalampiras2019} & Prt: 126; Rec: 920; Src: RSD & 4: Wheezes, Crackles, Wheezes+Crackles, and Normal & Acc: 50\%; Wheeze Sens: 65\%; Spec: 63\% \\
\cite{Demir2020} & Prt: 126; Rec: 920; Src: RSD & 4: Wheezes, Crackles, Wheezes+Crackles, and Normal & Acc: 66\%; Wheeze Prec: 55\%; Wheeze Sens; 43\%; Spec: 78\% \\\hline
\noalign{\vskip 1em}  
\end{tabular}%
}
Ref: reference; Prt: participants; Rec: recordings; Src: source; Priv: private; AUC: area under the Receiver Operating Characteristic curve; Acc: accuracy; Prec: precision; Sens: sensitivity; Spec: specificity; MCC: Matthews correlation coefficient; NA: not available
\end{table*}

With the onset of pattern recognition, several features and machine learning approaches have been proposed in the literature to develop automatic systems for wheeze classification. Pramono et al. \cite{Pramono2017a} identified the most common features and machine learning algorithms employed in the literature. These include Mel-frequency cepstral coefficients (MFCCs) \cite{Nakamura2016}, spectral features \cite{Bokov2016}, entropy \cite{Liu2016}, wavelet coefficients \cite{Ulukaya2017}, support vector machines (SVM) \cite{Lozano2016a}, artificial neural networks (ANN) \cite{Chamberlain2016}, rule-based models \cite{Taplidou2007} and logistic regression models \cite{Mendes2015}.

Over the years, several authors have reported excellent results on the wheeze classification task. A summary of selected works can be found in \autoref{tab:SOTA}. However, one crucial problem of this field has been its reliance on small or private data collections. Besides, public repositories that have been commonly used in the literature (e.g., R.A.L.E. \cite{Owens2002}) were designed for teaching, typically include a small number of adventitious respiratory sounds, and usually do not contain environmental noise. Therefore, we chose to perform the evaluation on the Respiratory Sound Dataset (RSD), the largest publicly available respiratory sound database, which is described in \autoref{sec:methods}.

\section{Materials and Methods}\label{sec:methods}

\subsection{Database}
The Respiratory Sound Database (RSD) is a publicly available database with 920 audio files containing a total of 5.5 hours of recordings acquired from 126 participants of all ages \cite{Rocha2019a}. The database main characteristics are described in \autoref{tab:demInfo}. The database contains audio samples that were collected independently by two research teams in two different countries and it is a challenging database, since the recordings contain several types of noises, background sounds and different sampling frequencies. It contains annotations of 1898 wheezes, which are found in 341 audio files. The training set contains 1173 wheezes distributed among 203 files, while the testing set includes 725 wheezes distributed among 138 files.

\begin{table}[b!]
\caption{Demographic information of database}
\label{tab:demInfo}
\begin{tabularx}{\columnwidth}{X|X}
\hline
\textbf{Number of recordings} & 920 \\
\textbf{Sampling frequency (number of recordings)} & 4 kHz (90); 10 kHz (6); 44.1 kHz (824) \\
\textbf{Bits per sample} & 16 \\
\textbf{Average recording duration} & 21.5 s \\
\textbf{Number of participants} & 126: 77 adults, 49 children \\
\textbf{Diagnosis} & COPD (64);   Healthy (26); URTI (14); Bronchiectasis (7); Bronchiolitis (6);  Pneumonia (6); LRTI (2); Asthma (1) \\
\textbf{Sex} & 79 males, 46 females (NA: 1) \\
\textbf{Age (mean $\pm$ standard deviation)} & 43.0 $\pm$ 32.2 years (NA: 1) \\
\textbf{Age of adult participants} & 67.6 $\pm$ 11.6 years (NA: 1) \\
\textbf{Age of child participants} & 4.8 $\pm$ 4.6 years \\
\textbf{BMI of adult participants} & 27.2 $\pm$ 5.4 kg m\textsuperscript{2} (NA: 2) \\
\textbf{Weight of child participants} & 21.4 $\pm$17.2 kg (NA: 5) \\
\textbf{Height of child participants} & 104.7 $\pm$ 30.8 cm (NA: 7)\\
\hline
\noalign{\vskip 1em}
\end{tabularx}
COPD: Chronic Obstructive Pulmonary Disease; LRTI: Lower Respiratory Tract Infection; NA: Not Available; URTI: Upper Respiratory Tract Infection.
\end{table}

\subsection{Random Event Generation}
To establish a comparison with other methods that were tested on this database, we mimicked “Experiment 2” of \cite{Rocha2019a} but focusing only on wheeze classification. As detailed in that experiment, we generated random events using a custom script. To simultaneously guarantee variation and reproducibility, the seed for the random number generator changed for each file but was predetermined. In this work, we employed two approaches to generate the random events. Both approaches generated the same number of events, as our goal was to study the impact of event duration on the algorithms’ performance. To keep the dataset classes' proportion similar to the original (40\% wheezes, 60\% other), we established a rule imposing a limit of one event per 5 s, i.e., the script generated at most four random events in a 20 s file, depending on the number of annotated wheezes. A total of 2910 random events were generated, 1781 in the training set and 1129 in the test set. In the first approach, following \cite{Rocha2019a}, events with a fixed duration of 150 ms were generated. In the second approach, we visually inspected the distribution of the annotated wheezes’ durations and found that a Burr distribution \cite{Burr1942} provided a good fit. The Burr distribution used to generate the events had probability density function:
\begin{equation}
f\left ( x\mid a,c,k  \right )=\frac{\frac{kc}{\alpha}\left ( \frac{n}{\alpha} \right )^{c-1}}{(1+\left ( \frac{n}{a} \right)^{c})^{k+1}}, x>0; \alpha>0; c>0; k>0
\end{equation}
with $\alpha=0.2266$\ , $c=4.1906$ and $k=0.3029$.

All the random events were then generated with a duration belonging to this distribution, ranging between 100 ms and 2 s.

\subsection{Preprocessing}
The audio files in RSD were recorded with different sampling rates. Therefore, we resampled every recording at 4000 Hz, the lowest sampling rate in the database. As the signal of interest was below 2000 Hz \cite{Bohadana2014a}, this was considered a good resolution for Fourier analysis.

\subsection{Spectrogram}
To generate the time-frequency representation of the audio events, the Short-Time Fourier Transform (STFT) was used.  The STFT is one of the most used tools in audio analysis and processing since it describes the evolution of the frequency components over time. The STFT representation (F) of a given discrete signal is given by \cite{Demir2020}:
\begin{equation}
F(n,\omega))=\sum_{i=-\infty}^{\infty}i\omega(n-i)e^{-j\omega}    
\end{equation}
where $\omega(i)$ is a window function centered at the instant $n$.

Since the database events had a wide range of durations, a maximum time for event was defined. Using the typical duration of an expiratory cycle, the maximum size per event was established as 2 seconds, and smaller events were centered and zero-padded. The database also contained annotated events longer than 2 s (77 events). For these cases, only the 2 first seconds were considered, as these bigger annotations were regarded as less precise. This stresses the need for a future review and correction of the database annotations. 

\begin{figure}
     \centering
     \begin{subfigure}[h]{0.49\columnwidth}
         \centering
         \includegraphics[width=\textwidth]{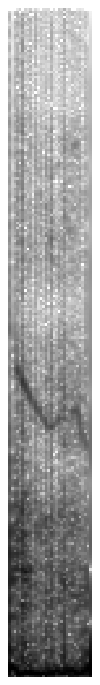}
         \caption{Wheeze Event}
     \end{subfigure}
     \begin{subfigure}[h]{0.49\columnwidth}
         \centering
         \includegraphics[width=\textwidth]{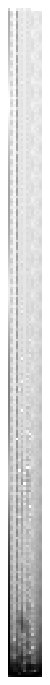}
         \caption{Random Event}
     \end{subfigure}
    \caption{Sample padded spectrograms (Left: Wheeze Event; Right: Random Event).}
    \label{fig:methods-spectrogram}
\end{figure}

Spectrograms with a 128 ms hamming window and 75\% of overlap were generated. The spectrograms were normalized between 0 and 1. The spectrogram images used by the CNN had 59 by 257 pixels. In \autoref{fig:methods-spectrogram}, it is possible to observe an example of two representative events. 

\subsection{Feature Extraction}
In this step, 47 features were extracted from each frame of the spectrogram using the MIR Toolbox \cite{Lartillot2007}. Most of these features were used in \cite{Mendes2015} for wheeze classification and are commonly used in the field of music information research. \autoref{tab:features} provides a small description of all the features. Please note that the selected settings were based on previous research and it was not possible to validate all the frame sizes and other extraction parameters for this study. For each event, five statistics of each feature were calculated: mean, standard deviation, median, minimum value, and maximum value. Therefore, the total number of features fed to the classifiers was 235.

\begin{table}[t!]
\centering
\caption{Features description.}
\label{tab:features}
\begin{tabularx}{\columnwidth}{p{2.6cm}|X}
\hline
\textbf{Feature} & \textbf{Description} \\\hline
Spectral Centroid & Center of mass of the spectral distribution \\
Spectral Spread & Variance of the spectral distribution \\
Zero-crossing Rate & Waveform sign-change rate \\
Spectral Entropy & Estimation of the complexity of the spectrum \\
Spectral Flatness & Estimation of the noisiness of a spectrum \\
Spectral Roughness & Estimation of the sensory dissonance \\
Spectral Irregularity & Estimation of the spectral peaks’ variability \\
Spectral Flux & Euclidean distance between the spectrum of successive frames \\
Spectral Brightness & Amount of energy above \{100,200,400,800\} Hz \\
Spectral Rolloff & Frequency such that \{95,75,25,5\}\% of the total energy is contained below it \\
MFCC & 13 Mel-frequency cepstral coefficients \\
Delta-MFCC & 1st-order temporal differentiation of the MFCCs \\
Chromagram Centroid & Tonal centroid \\
Chromagram Peak & Tonal peak \\
Pitch & Fundamental frequency estimation \\
Voicing & Presence of fundamental frequency \\
Inharmonicity & Partials non-multiple of fundamental frequency \\\hline
\end{tabularx}
\end{table}

\subsection{Classifiers}
To classify the samples as wheezes or other we used several machine learning algorithms: linear discriminant analysis (LDA), Linear SVM (SVMlin), Gaussian SVM (SVMrbf), boosted trees (Boost), and a CNN (using the spectrogram images as input). All the classifiers were trained 10 times with different seeds and their parameters were optimized on a validation set containing 25\% of the training set. The models with the best parameters were then applied to the testing set.
Bayes optimization was used to optimize the following parameters of each traditional machine learning algorithm: delta and gamma for LDA; box constraint and kernel scale for SVM; method, number of learning cycles (NLearn), and learning rate (LearnRate) for Boost. The best parameters for the best run of each system are listed in \autoref{tab:tableClassifiers}.

\begin{table}[b!]
\centering
\caption{Classifiers' hyperparameters (best models).}
\label{tab:tableClassifiers}
\begin{tabularx}{\columnwidth}{p{1.1cm}|X|X}
\hline
\textbf{Algorithm} & \textbf{Parameters (Fixed Durations)} & \textbf{Parameters (Variable Durations)} \\ \hline
LDA & Delta: 0.00088684; Gamma: 0.001 & Delta: 0.0104; Gamma: 0.0405 \\ \hline
SVMlin & BoxConstraint: 931.133; KernelScale: 425.5836 & BoxConstraint: 9.6309; KernelScale: 923.6736 \\ \hline
SVMrbf & BoxConstraint: 86.0544; KernelScale: 990.1481 & BoxConstraint: 27.2645; KernelScale: 999.2519 \\ \hline
Boost & Method: Logitboost; NLearn: 186; LearnRate: 0.2582 & Method: LogitBoost; NLearn: 264; LearnRate: 0.6745 \\ \hline
\end{tabularx}
\end{table}

To develop the CNN model, a base architecture was defined and the parameters of each layer were then optimized using a grid search approach. The optimized parameters were the following ones: Convolution size; Number of convolutional filters; Dropout rate; Pooling size; Size of the fully connected layer. The models were trained with a maximum number of 15 epochs, a mini batch size of 128, Adam optimization (adaptive moment estimation), and 10\% of the training dataset was used for validation during the training phase. The best models’ architectures for both experiments are described in \autoref{tab:cnns}.

A system consisting of all the features and a logistic regression model was used as a baseline for comparison.

\begin{table}[t!]
\centering
\caption{CNN's architecture.}
\label{tab:cnns}
\begin{tabularx}{\columnwidth}{X|X|X|X|X}
\hline
\textbf{Layer} & \textbf{Activations (Fixed Durations)} & \textbf{Parameters (Fixed Durations)} & \textbf{Activations (Variable Durations)} & \textbf{Parameters (Variable Durations)} \\ \hline
Input & 257 x 59 x 1 & - & 257 x 59 x 1 & - \\ \hline
2D Convolution & 251 x 53 x 64 & Convolution Size-7; Stride-1; Filters-64 & 253 x 55 x 32 & Convolution Size-5; Stride-1; Filters-32 \\ \hline
Batch normalization & 251 x 53 x 64 & - & 253 x 55 x 32 & - \\ \hline
ReLU & 251 x 53 x 64 & - & 253 x 55 x 32 & - \\ \hline
Max Pooling & 250 x 52 x 64 & Pooling Size-2; Stride-1 & 250 x 55 x 32 & Pooling Size-4; Stride-1 \\ \hline
Dropout & 250 x 52 x 64 & 50\% & 250 x 55 x 32 & 50\% \\ \hline
Fully Connected & 1 x 1 x 10 & Size-10 & 1 x 1 x 20 & Size-20 \\ \hline
Dropout & 1 x 1 x 10 & 50\% & 1 x 1 x 20 & 50\% \\ \hline
Fully Connected & 1 x 1 x 2 & Size-2 & 1 x 1 x 2 & Size-2 \\ \hline
Softmax & 1 x 1 x 2 & - & 1 x 1 x 2 & - \\ \hline
\end{tabularx}
\end{table}

\subsection{Evaluation Metrics}
To evaluate the algorithms’ performance, we used the following measures:
\begin{equation}
Accuracy(Acc)=\frac{(TP+TN)}{(TP+TN+FP+FN)}
\end{equation}
\begin{equation}
Precision(Prec)=\frac{TP}{(TP+FP)}
\end{equation}
\begin{equation}
Sensitivity(Sens)=\frac{TP}{(TP+FN)}
\end{equation}
\begin{equation}
F1 Score(F1)=\frac{(2 \times Prec\times Sens)}{(Prec + Sens)}
\end{equation}
\begin{equation}
Specificity(Spec)=\frac{TN}{(TN+FP)}
\end{equation}
\begin{equation}
\resizebox{\columnwidth}{!}{$Matthews Corr Coef (MCC)= \frac{((TP \times TN)-(FP \times FN))}{\sqrt{((TP+FP)(TP+FN)(TN+FP)(TN+FN))}}$}
\end{equation}

\noindent where TP (True Positives) are wheeze events that are correctly classified; TN (True Negatives) are other events that are correctly classified; FP (False Positives) are other events that are incorrectly classified as wheezes; FN (False Negatives) are wheeze events that are incorrectly classified as other.

\section{Evaluation}\label{sec:evaluation}
In this section, we analyze the performance of the algorithms on two experiments corresponding to the two approaches for the generation of random events. We focus especially on MCC, as this is the only evaluation measure that simultaneously considers TP, TN, FP, and FN. To analyze statistical significance, we performed the right-tailed Wilcoxon sign rank test, for a significance level of $\alpha$ = 1\%, against the baseline. To adjust for multiple comparisons, we applied Bonferroni correction to the alpha value. \autoref{tab:TrainTestData} provides details about the training and testing sets.

\begin{table}[b!]
\centering
\caption{Training and test dataset classes distribution.}
\label{tab:TrainTestData}
\begin{tabularx}{\columnwidth}{c|c|c|c}
\hline
\textbf{} & \textbf{Training set} & \textbf{Testing set} & \textbf{Total} \\ \hline
\textbf{Number of wheezes} & 1173 & 725 & 1898 \\ \hline
\textbf{Number of random events} & 1781 & 1129 & 2910 \\ \hline
\textbf{Total} & 2954 & 1854 & 4808 \\ \hline
\end{tabularx}
\end{table}

\subsection{Fixed Durations}
\autoref{tab:resultsFixed} displays the results achieved by all the combinations of classifiers and feature sets on the fixed durations (FD) test set. \autoref{fig:mccBarPlot} shows the MCC results with the corresponding error bars.

\begin{table*}[t!]
\centering
\caption{Performance results on FD test set (* statistically significant difference ($\alpha=0.01$)) (mean $\pm$ standard deviation)}
\label{tab:resultsFixed}
\resizebox{0.9\textwidth}{!}{%
\begin{tabular}{c|c|c|c|c|c|c}
\hline
 & \textbf{Accuracy} & \textbf{Precision} & \textbf{Sensitivity} & \textbf{F1} & \textbf{Specificity} & \textbf{MCC} \\ \hline
Baseline & 90.5$\pm$0 & 96$\pm$0 & 79$\pm$0 & 86.7$\pm$0 & 97.9$\pm$0 & 80.3$\pm$0 \\ \hline
LDA & 89.1$\pm$0.4 & 94.2$\pm$0.5 & 76.8$\pm$1.3 & 84.6$\pm$0.7 & 97$\pm$0.3 & 77.2$\pm$0.8 \\ \hline
SVMlin & 87.7$\pm$2.5 & 91.1$\pm$7.4 & 76.7$\pm$3.2 & 83$\pm$2.5 & 94.7$\pm$5.3 & 74.3$\pm$5.1 \\ \hline
SVMrbf & 88.4$\pm$0.5 & 90.2$\pm$0.8 & 78.8$\pm$1.8 & 84.1$\pm$0.9 & 94.5$\pm$0.6 & 75.4$\pm$1.1 \\ \hline
Boost & 90.6$\pm$0.7 & 91.7$\pm$2.4 & 83.6$\pm$1.6 * & 87.4$\pm$0.8 & 95$\pm$1.6 & 80.1$\pm$1.4 \\ \hline
CNN & 96$\pm$2.2 * & 91.9$\pm$4.3 & 98.1$\pm$5.1 * & 94.7$\pm$2.8 * & 95.1$\pm$2.4 & 91.8$\pm$4.4 * \\ \hline
\end{tabular}%
}
\end{table*}

As mentioned in \autoref{sec:background}, the best system’s wheeze Sens and Spec in \cite{Rocha2019a} were 78.6\% and 90.5\%, respectively. With Sens and Spec of 79\% and 97.9\%, respectively, our baseline's performance was on par with the state-of-the-art. In addition, it achieved better results than the other traditional classifiers (MCC of 80.3\%). Nevertheless, the CNN achieved even better results, reaching an MCC of 91.8\%. Considering the MCC, only the CNN performed significantly better than the baseline.

\subsection{Variable Durations}
\autoref{tab:resultsVariable} displays the results achieved by all the combinations of classifiers and feature sets on the variable durations (VD) test set. \autoref{fig:mccBarPlot} shows the MCC results with the corresponding error bars.

\begin{table*}[t!]
\centering
\caption{Performance results on VD test set (* statistically significant difference ($\alpha=0.01$)) (mean $\pm$ standard deviation)}
\label{tab:resultsVariable}
\resizebox{0.9\textwidth}{!}{%
\begin{tabular}{c|c|c|c|c|c|c}
\hline
 & \textbf{Accuracy} & \textbf{Precision} & \textbf{Sensitivity} & \textbf{F1} & \textbf{Specificity} & \textbf{MCC} \\ \hline
Baseline & 67.2$\pm$0 & 58.4$\pm$0 & 55.9$\pm$0 & 57.1$\pm$0 & 74.5$\pm$0 & 30.6$\pm$0 \\ \hline
LDA & 66.2$\pm$0.2 & 57.8$\pm$0.4 & 50.2$\pm$1.0 & 53.7$\pm$0.6 & 76.5$\pm$0.6 * & 27.5$\pm$0.5 \\ \hline
SVMlin & 66.7$\pm$0.2 & 60.5$\pm$0.7 * & 42.7$\pm$1.2 & 50.1$\pm$0.7 & 82.1$\pm$0.9 * & 27.1$\pm$0.5 \\ \hline
SVMrbf & 62.9$\pm$0.7 & 53.2$\pm$1.0 & 43.6$\pm$1.3 & 47.9$\pm$1.1 & 75.4$\pm$0.8 & 19.8$\pm$1.4 \\ \hline
Boost & 67.8$\pm$1.3 & 59.6$\pm$2.1 & 55.3$\pm$4.3 & 57.2$\pm$2.4 & 75.9$\pm$3.1 & 31.7$\pm$2.7 \\ \hline
CNN & 60.3$\pm$5.2 & 67.5$\pm$20.4 & 51$\pm$4.9 & 55.9$\pm$7.4 & 74.7$\pm$5.4 & 24.4$\pm$4.5 \\ \hline
\end{tabular}%
}
\end{table*}

The decline in performance is quite noticeable. The best classifier in this experiment, Boost, reached an MCC of 31.7\%. However, none of the optimized algorithms significantly outperformed the baseline (MCC of 30.6\%).

\subsection{Variance in performance}
\begin{figure}[b!]
\centering
\includegraphics[width=0.5\textwidth]{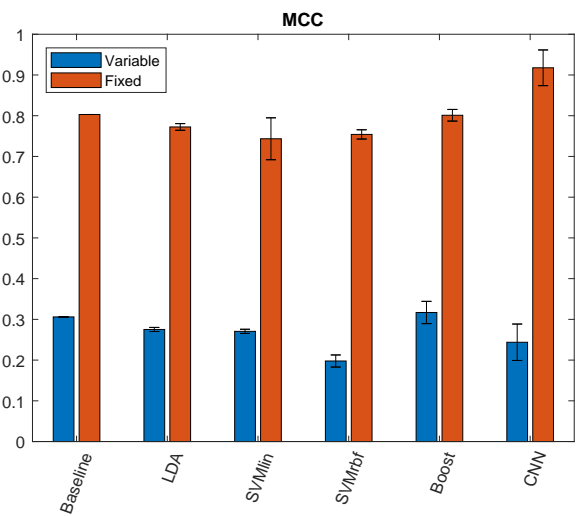}
\caption{MCC values for 10 runs in both experiments (mean and standard deviation).}
\label{fig:mccBarPlot}
\end{figure}

\begin{figure*}[t!]
     \centering
     \begin{subfigure}[h]{0.43\textwidth}
        \centering
        \includegraphics[width=\columnwidth]{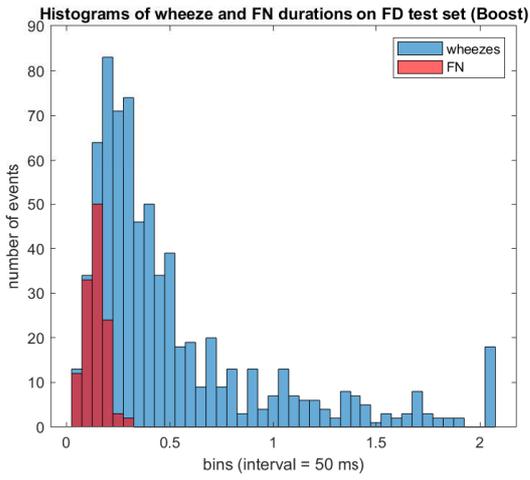}
        \caption{FD test set (best run – Boost).}
        \label{fig:fnPlotFixedClassifier}
     \end{subfigure}
     \begin{subfigure}[h]{0.43\textwidth}
        \includegraphics[width=\columnwidth]{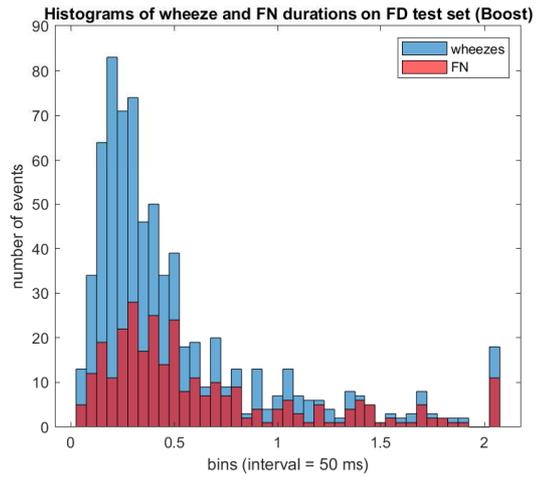}
        \caption{VD test set (best run – Boost).}
        \label{fig:fnPlotVariableClassifier}
     \end{subfigure}
     
     \begin{subfigure}[h]{0.43\textwidth}
        \centering
        \includegraphics[width=\columnwidth]{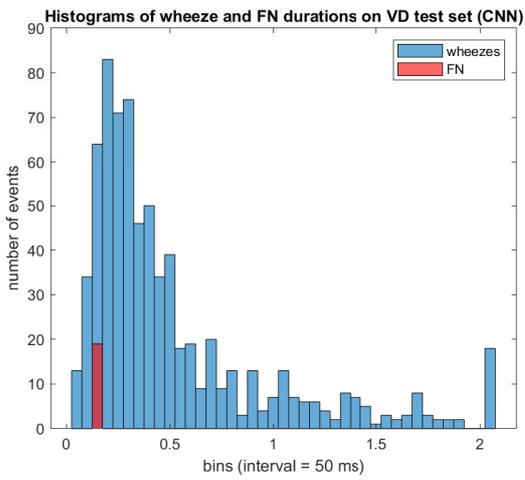}
        \caption{FD test set (best run – CNN).}
        \label{fig:fnPlotFixedCNN}
     \end{subfigure}
     \begin{subfigure}[h]{0.43\textwidth}
        \includegraphics[width=\columnwidth]{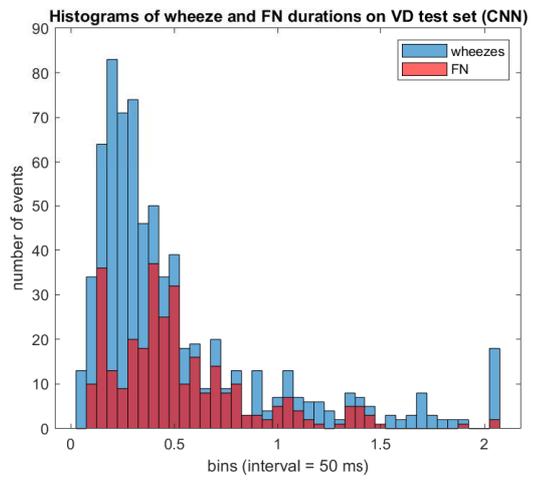}
        \caption{VD test set (best run – CNN).}
        \label{fig:fnPlotVariableCNN}
     \end{subfigure}
     \caption{Histograms of wheezes and FN durations.}
\end{figure*}

The substantial variance in performance between experiments might indicate that the generation of the random events with fixed durations introduced considerable bias. \autoref{fig:mccBarPlot} shows the MCC results for each experiment with the corresponding error bars. The classifiers could be implicitly learning to identify the duration of the events. To analyze this bias, we can plot the histogram of event durations of the errors the best classifiers made. The best run of the Boost model yielded 124 FN. As displayed in \autoref{fig:fnPlotFixedClassifier}, all the misclassified wheezes had short durations of less than 325 ms. When classifying wheezes against random events of variable duration, the distribution of FN durations was quite different, encompassing all durations, as shown in \autoref{fig:fnPlotVariableClassifier}. When using the CNN to perform the same classification, the effect of implicitly learning the duration of events was even more noticeable, as can be seen when comparing \autoref{fig:fnPlotFixedCNN} and \autoref{fig:fnPlotVariableCNN}. All the misclassified wheezes in the fixed durations test set belonged to the same histogram bin, comprising durations between 125 ms and 175 ms. When considering the variable durations, the CNN model’s FN distribution also covered the full range of durations. To improve visualization, all the annotated wheezes longer than 2 s were equalized and assigned to the same histogram bin.

\begin{figure*}[t!]
\centering
\includegraphics[width=0.9\textwidth]{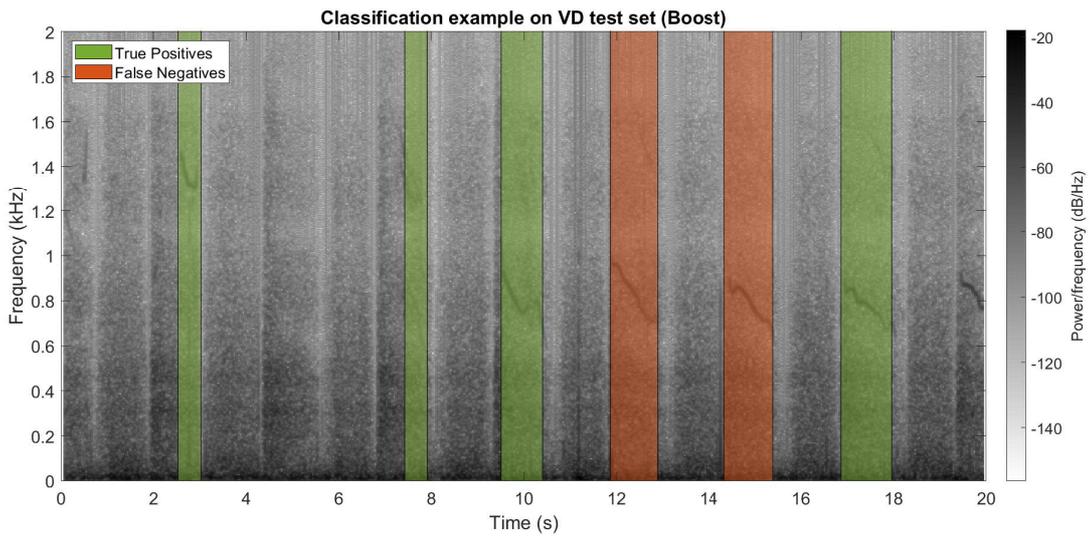}
\caption{Classification example on VD test set (file: 133\_3p4\_Tc\_mc\_AKGC417L)}
\label{fig:ExampleClass}
\end{figure*}

\autoref{fig:ExampleClass} illustrates how the same algorithm attributed different classes to the same events depending on how the non-wheeze class was designed. On the FD experiment, all annotated wheezes in this sound file were correctly classified, while on the VD experiment, two out of six annotated wheezes were incorrectly classified. Through visual inspection of the spectrogram alone, it is not possible to understand why the 3\textsuperscript{rd} and 6\textsuperscript{th} wheezes were correctly classified while the 4\textsuperscript{th} and 5\textsuperscript{th} wheezes were not, as they seem to show similar spectrogram signatures. This figure also reveals potential annotation inaccuracies, as the spectrogram suggests a possible wheeze that was not annotated between 19.4 s and 20 s.

\section{Discussion and Conclusion}\label{sec:discussion}
In this work, we proposed and evaluated several wheeze classification systems, and demonstrated how event duration can have a significant impact on automatic wheeze classification. As the algorithms’ performance presented in the previous section show, methods that seem to achieve promising results can fail if we change the way the non-wheeze class is designed. This can happen even if the dataset where the systems are evaluated does not change. It is important to consider how data are used to train, validate, and test a classification algorithm. Our goal was not to question state-of-the-art results in wheeze classification, but to emphasize the critical role of experimental design in the evaluation of the algorithms' performance. As asserted by Sturm \cite{Sturm2014}, when the performance of an artificial system appears to support the claim that the system is addressing a complex human task (e.g., classify sounds as wheezes), the default position (null hypothesis) should be that the system is not actually addressing the problem it appears to be solving; in this case, a system that claims to distinguish a wheeze from a random event is implicitly learning an irrelevant characteristic of the dataset, i.e., event duration. To determine if a system is relevant, we need to understand the extent to which the characteristics it is extracting from the signal are confounded with the ground truth, i.e., the expert's annotations. Besides, investigations like this are only possible if algorithm evaluation is performed on open access datasets.

Nevertheless, it is important to reiterate the limitations of this database, which may have influenced the performance of the evaluated systems. As pointed out in \cite{Rocha2019a}, these include the lack of healthy adult participants and the absence of gold standard annotations, i.e., annotations from multiple annotators. A future update of the database should also check for possible errors.

Wheeze classification is a complex task that is not yet solved, despite the claims made in the literature. It may be particularly hard when algorithms are evaluated on challenging datasets, such as the RSD. Though significant work has been developed to classify wheezes, none has been widely accepted \cite{marquesJacom2018}. While CNNs have become state-of-the-art solutions in several tasks \cite{Bardou2018}, a simple CNN with no preprocessing was not enough to tackle this problem. Therefore, accelerating the development of machine learning algorithms is critical to the future of respiratory sounds analysis.

\section*{Acknowledgment}
This work was partially supported by: Fundação para a Ciência e Tecnologia (FCT) PhD scholarship SFRH/BD/135686/2018; WELMO project, co-funded by the Horizon 2020 Framework Programme of the European Union under grant agreement 825572; Fundo Europeu de Desenvolvimento Regional (FEDER) through Programa Operacional Competitividade e Internacionalização (COMPETE) and FCT under the project UID/BIM/04501/2013 and POCI-01-0145-FEDER-007628—iBiMED.

\bibliography{ArticleSensors.bib}
\bibliographystyle{ieeetr}

\end{document}